# The Wisdom of Citing Scientists

Lutz Bornmann* & Werner Marx**

*First author and corresponding author:

Division for Science and Innovation Studies,

Administrative Headquarters of the Max Planck Society,

Hofgartenstr. 8,

80539 Munich, Germany.

E-mail: bornmann@gv.mpg.de

**Second author:

Max Planck Institute for Solid State Research,

Heisenbergstraße 1,

D-70569 Stuttgart

**Abstract:**

This Brief Communication discusses the benefits of citation analysis in research evaluation based on Galton's "Wisdom of Crowds" (1907). Citations are based on the assessment of many which is why they can be ascribed a certain amount of accuracy. However, we show that citations are incomplete assessments and that one cannot assume that a high number of citations correlate with a high level of usefulness. Only when one knows that a rarely cited paper has been widely read is it possible to say – strictly speaking – that it was obviously of little use for further research. Using a comparison with 'like' data, we try to determine that cited reference analysis allows a more meaningful analysis of bibliometric data than times-cited analysis.

**Introduction**

The evaluation of research is the backbone of modern science: it would not be possible to determine whether a piece of research is of high or low quality without a review by a researcher's peers (Bornmann, 2011). The importance of the evaluation of research manifests itself firstly in the fact that since around the middle of the 17th century the development of modern science in society has been closely associated with the establishment of the peer review process (de Bellis, 2009). It seems that modern science requires a formal process with which to evaluate scientific work so that knowledge can continue to progress. The importance of the evaluation of research is also underlined by Robert K. Merton's inclusion of "organised scepticism" as a key norm in the ethos of science (Merton, 1973): "Organized skepticism involves a latent questioning of certain bases of established routine, authority, vested procedures and the realm of the 'sacred' generally" (p. 264). Along with communalism (making knowledge public), universalism (recognition of knowledge irrespective of its source) and disinterestedness (not working for profit), organised scepticism, according to Merton (1973), should be a guiding principle for researchers.

While peer review has been used exclusively to assess research in almost every scientific discipline, since the mid-1980s the use of indicators to evaluate research has become increasingly important – at least in the natural sciences. Bibliometrics in particular has become a focus of interest as an indicator-based method of assessment. Although the strengths of each approach mean that bibliometrics is primarily suited to the assessment of large units (such as institutions or countries) and peer review to evaluation on a smaller scale (such as research proposals or manuscripts), informed peer review is seen as the ideal way forward for evaluation in research (Bornmann, 2011). With this approach, results based on evaluations with indicators form a solid basis on which experts in the peer review process can take their decisions. Bibliometrics (or its most important tool, citation analysis) is primarily of benefit when evaluating large units because it is based on the judgement of a large number of scientists. A published work contains research results which (1) can be accessed by any other member of the scientific community, (2) these scientists can use for their own research and (3) they can cite in their own publications. A large number of citations of a publication indicates that it has been interesting and useful for a large number of researchers (Bornmann & Marx, 2013a).

In this Brief Communications, we point out (1) the high accuracy level of citations, because citations are based on the opinions of a large number of scientists ("Wisdom of Crowds"). Since citations are count data and not ordinal data, we argue (2) that citations are also an incomplete judgement. As count data, we think that (3) citations, in the form of cited references, say more about the citing unit than citations about the cited unit. The cited reference analysis should therefore be preferred to the times cited analysis.

**The Wisdom of Crowds**

Citations measure an aspect of scientific quality – the impact of publications (van Raan, 1996). Martin and Irvine (1983) distinguish between this aspect ("the 'impact' of a publication describes its *actual* influence on surrounding research activities at a given time," p. 70) and 'importance' ("the influence on the advance of scientific knowledge," p. 70) and 'quality' ("how well the research has been done," p. 70). They consider the impact as the most important indicator of the significance of a publication for scientific activities. Cole (1992) sees citations as a valid indicator of quality, as they correlate with other quality indicators (see also Bornmann, 2011; Smith & Eysenck, 2002).

The benefit of citation analysis for the evaluation of research is based on what Galton (1907) at the beginning of the 20th century called the "wisdom of crowds". Galton (1907)'s reports on the results "of a contest at an English livestock show where contestants were asked to guess the weight of a publicly displayed ox. After sorting the 787 entries by weight, Galton found that the median estimate of 1,207 pounds differed from the true weight of 1,198 pounds by less than 1%" (Herron, 2012, p. 2278). When a large number of people make a judgement or an estimate, we can expect it to be valid. Citation analysis is also based on the judgement of many people: the scientific community cites a paper when it has turned out to be useful, or not when it is less useful. Galton (1907)'s premise on the validity of such crowd judgements forms the basis for Surowiecki (2004)'s book "The Wisdom of Crowds". However Surowiecki (2004) argues that one should not assume that all crowds are wise. Crowds are only wise and their judgement only achieves a high level of accuracy when the individual judgements are based on opinions which were delivered independently. If we were to apply this proviso of Surowiecki (2004)'s to bibliometrics, we could also assume that citation analysis allows a high level of accuracy: firstly, the citations are based on the opinions of a large number of scientists. Secondly, we can assume that most citations are based on the independent opinion of the citing researcher. There are just a few cases in which publishers or reviewers have asked that citations be made, that they be discussed with colleagues or not used for immaterial considerations.

However, unlike assessments of the weight of an ox, citations represent an incomplete judgement. A citation denotes the usefulness of a paper, but it is not possible to use citations to express a graduated assessment. Citations are count data (there is or there is not a citation) and not ordinal data which would convey better or worse, important or unimportant. Furthermore, a citing researcher only expresses which publications he or she has found useful and does not indicate which publications were irrelevant to the writing of the paper. We cannot, therefore, automatically assume when a publication is not cited that it is not important to research. We could only assume that an uncited publication was not particularly useful for research if we knew that it had been widely read. On the other hand, a publication that has gone largely unnoticed has few opportunities to be cited – irrespective of its quality.

Nevertheless, we would not like to interpret the classification of citations as count data and the therefore accompanying disadvantages in the evaluation of publications as meaning that citations have no use in the "wisdom of peers" research evaluation. As there are two perspectives to citation analysis – citing and cited – we would like to posit the provocative claim in this paper that citations make a powerful statement particularly in one of these perspectives – the cited perspective. Citations, in the form of cited references, say more about the citing unit than citations about the cited unit.

To illustrate this provocative claim, we want to compare citations with the 'like' buttons which are used in social network services, blogs, Internet forums and so on to explain the preference for the cited perspective in citation analysis. Users click 'like' buttons to indicate that they approve of content (such as a book or a piece of music). Similarly to the times-cited data provided by Thomson Reuters in the Web of Science (WoS), Internet services give the number of users who like certain content on the basis of this information. This analogy between 'likes' and citations does not focus on the cognitive substance, but the measurement category of the data. Whereas the measurement category is very similar (both are count data), the cognitive substance is very different. As concept symbols (Small, 1978) citations have a significant greater substance than 'likes' which express a certain preference only.

The quantity of users of a product expressed by 'likes' says little about how much a content is generally approved of. (1) As count data, 'likes' do not permit a graduated assessment. (2) Because it is not known how many users are familiar with the content, but have not evaluated it, knowing the number of users who 'like' it tells us little. It is only the proportion of users who like the content among all the users who have delivered an assessment which would allow a comparison of the popularity of different content. Unlike the citation data in bibliometrics, the 'like' data is used as a rule not to assess the content, but to assess the users who have made a statement about certain content. The 'like' data is therefore evaluated for the person who has provided information and not the content which is being evaluated. The user-specific statements on various contents can be used to compile a user profile which provides information about a user's preferences. This data is of great significance for targeted, individualised advertising.

**Cited reference analysis as a tool for research evaluation**

Nor, as a rule, is it known for an academic publication how many scientists have read it, but have not considered it very useful (and have not cited it for this reason). The 'times-cited' information is therefore only of limited significance. Reference sets are used to achieve standardisation by comparing citation impact across different fields (Bornmann & Marx, 2013a) but the information about how many scientists have not found the publication very useful is not available. Similarly to the 'like' button, it would also be possible to argue that citations might be better used to characterise the citing unit.

References (citations) consist fundamentally of three pieces of information: the author(s) of the cited paper (author name), the journal in which the cited paper appears (including information such as volume, issue and pages) and its year of publication. A reference analysis can refer to any one of these pieces of information. More data can be included: for example, it is possible to find out for each journal which publishing house it is issued by, whether it is an open access journal or whether it operates using the peer review process. For researchers, it is possible to determine which journals or publishing houses they refer to in their publications and whether they use primarily quality-assured papers in their work (that is, papers from journals which operate a peer review process). Further information about the affiliation of the authors can be used for bibliometric analysis at institutional or country level. The analysis of the publication years could reveal the extent to which researchers read the current literature in their field, and whether they refer back to the historical roots of their discipline.

However cited reference analysis can be used for more than compiling a profile of the citing scientist; it can also be used for compiling profiles of other citing units. These might be research groups, institutions or countries, or even disciplines or researched subjects. Bornmann and Marx (2013b), for example, have carried out a cited reference analysis on the subject of research into Aspirin®. First, they selected all the papers dealing with Aspirin® as a specific research topic. Then they extracted all the cited references from this field-specific publication set and analysed which papers, authors and journals were cited most often. In other words: they categorised on the basis of the cited references rather than on the cited papers in this specific field. Cited reference analysis characterised research on Aspirin® by identifying, for example, the journals preferred by the citing scientists.

To illustrate the difference between traditional citation analysis (times-cited) and reference analysis (cited reference counts) we would like to use a dynamically growing area in which currently very intensive research is being undertaken: research associated with graphene, an allotrope of carbon (Geim & Kim, 2008; Geim & Novoselov, 2007). The papers published in 2010 with the word 'graphene' in the title were selected in the Science Citation Index (SCI). The SCI is made available for such analyses by the data provider STN International. Table 1 shows the bibliographic data of the publications on graphene research from 2010 from a forward view (times-cited) and a backward view (cited reference counts). The most-highly cited papers <u>from</u> graphene research are shown on the left-hand side and the papers most-highly cited <u>in</u> graphene research are shown on the right-hand side. While the right-hand side shows which publications have been particularly significant for research associated with graphene, we only see from the data on the left-hand side that a number of graphene publications were significant for the research in general. However, we do not know for which research the publications were so significant.

Cited reference analyses are uncommon in citation impact measurement but they are used for techniques such as bibliographic coupling (Kessler, 1963) or citing-side journal mapping (Leydesdorff, 1994). One of the rare examples are the cited reference analyses for the "science and engineering indicators" (National Science Board, 2012): the share of world citations is shown for a series of selected countries and specific times. Another example is the study by Bornmann, de Moya-Anegón, and Leydesdorff (2010) on the "Ortega Hypothesis". This study shows that high-impact research uses earlier high-impact research more than medium-impact research does.

The usefulness of the cited reference perspective can be justified by the fact that many bibliometric studies (and beyond) contain a note indicating that the citation impact of a paper, scientist or group of scientists should be or is measured in a particular field. For example, "the result is the identification of high performers within a given scientific field" (Froghi et al., 2012, p. 321). "Ideally, a measure would reflect an individual's relative contribution within his or her field" (Kreiman & Maunsell, 2011). "That is, an account of the number of citations received by a scholar in articles published by his or her field colleagues" (Di Vaio, Waldenström, & Weisdorf, 2012, p. 92). The well-known philosopher of science and American historian Thomas S. Kuhn formulated: "For a scientist, the solution of a difficult conceptual or instrumental puzzle is a principal goal. His success in that endeavour is rewarded through recognition by other members of his professional group and by them alone" (Kuhn, 1970, p. 21). However, with the standard times cited analysis, it is not just the citation impact on a specific field that is measured, but the impact in the whole scientific community.

**Discussion**

This Brief Communication discusses the benefits of citation analysis in research evaluation based on Galton's "Wisdom of Crowds" (1907). Citations are based on the assessment of many which is why they can be ascribed a certain amount of accuracy. However, we have also shown that citations are incomplete assessments and that one cannot assume that a high number of citations correlate with a high level of usefulness. Only when one knows that a rarely cited paper has been widely read is it possible to say – strictly speaking – that it was obviously of little use for further research. Using a comparison with 'like' data, we have tried to determine that cited reference analysis allows a more meaningful analysis of bibliometric data than times-cited analysis.

For example a cited reference analysis for one scientist could provide information about the extent to which he or she refers to older or more recent publications, which journals he or she reads most (and then cites), the research is based, which theoretical approaches to research he or she prefers and the quality of the cited paper. This latter could provide information about whether the scientist is able to identify high-quality publications and include them in his or her own work.

An evaluation of research which does not use citations is unthinkable in the natural sciences nowadays (Bornmann et al., in press; Bornmann & Marx, 2013c). The great popularity of indicators such as the h index and the journal impact factor (although not the most appropriate bibliometric indicators) are strong evidence of this. The importance of citations for research evaluation purposes has been challenged neither by the existence of negative citations, self-citations, o gift-citations (Bornmann & Daniel, 2008) and of the well-known phenomenon 'obliteration by incorporation' (McCain, 2011; Merton, 1968), nor by recent discussions on the correlations between citation patterns and the status of findings in science as 'creative,' 'mainstream,' 'contested,' and 'ignored' (Heinze, 2013).

Even if many studies suggest that not only the content of scientific work, but also other, in part non-scientific factors play a role in citing behaviour, one should not conclude that citations are an inappropriate indicator of impact of research. According to van Raan (2005) the process of citation certainly does not provide "an 'ideal' monitor on scientific performance. This is particularly the case at a statistically low aggregation level, e.g. the individual researcher. There is, however, sufficient evidence that these reference motives are not so different or 'randomly given' to such an extent that the phenomenon of citation would lose its role as a reliable measure of impact. Therefore, application of citation analysis to the entire work, the 'oeuvre' of *a group of researchers as a whole over a longer period of time* (author's emphasis), does yield in many situations a strong indicator of scientific performance" (pp. 134-135).

With this Brief Communication, it is not our intention to cast further doubt on the use of citation analysis for research evaluation; we merely wish to posit – somewhat provocatively – the statement that given the limited significance of times-cited data, in many cases a cited reference analysis would make more sense. We would be pleased if putting this thesis forward has stimulated lively discussion amongst bibliometricians.

Table 1: Times-cited and cited reference analysis using the example graphene research in 2010.

| Times cited | Most-cited paper from graphene research in 2010 | Cited reference counts | Papers most-cited in graphene research in 2010 |
|---|---|---|---|
| 1054 | BAE S, NAT NANOTECHNOL 2010 V5 P574 | 1272 | NOVOSELOV K S, SCIENCE 2004 V306 P666 |
| 942 | DREYER D R, CHEM SOC REV 2010 V39 P228 | 928 | GEIM A K, NAT MATER 2007 V6 P183 |
| 637 | ZHU Y W, ADV MATER 2010 V22 P3906 | 708 | NOVOSELOV K S, NATURE 2005 V438 P197 |
| 633 | ALLEN M J, CHEM REV 2010 V110 P132 | 640 | CASTRONETO A H, REV MOD PHYS 2009 V81 P109 |
| 584 | LIN Y M, SCIENCE 2010 V327 P662 | 629 | ZHANG Y B, NATURE 2005 V438 P201 |
| 579 | BONACCORSO F, NAT PHOTONICS 2010 V4 P611 | 383 | STANKOVICH S, NATURE 2006 V442 P282 |
| 571 | SCHWIERZ F, NAT NANOTECHNOL 2010 V5 P487 | 351 | BERGER C, SCIENCE 2006 V312 P1191 |
| 450 | DEAN C R, NAT NANOTECHNOL 2010 V5 P722 | 342 | HUMMERS W S, J AM CHEM SOC 1958 V80 P1339 |
| 401 | QU L T, ACS NANO 2010 V4 P1321 | 311 | KIM K S, NATURE 2009 V457 P706 |
| 374 | CAI J M, NATURE 2010 V466 P470 | 309 | FERRARI A C, PHYS REV LETT 2006 V97 P187401 |
| 371 | ZHANG H, ACS NANO 2010 V4 P380 | 307 | GEIM A K, SCIENCE 2009 V324 P1530 |
| 369 | WU Z S, ACS NANO 2010 V4 P3187 | 294 | STANKOVICH S, CARBON 2007 V45 P1558 |

Note. Source: SCI from STN International